\documentclass[american,aps,pra,reprint, superscriptaddress]{revtex4-1}
\usepackage[unicode=true,pdfusetitle, bookmarks=true,bookmarksnumbered=false,bookmarksopen=false, breaklinks=false,pdfborder={0 0 0},backref=false,colorlinks=false] {hyperref}
\hypersetup{ colorlinks,linkcolor=myurlcolor,citecolor=myurlcolor,urlcolor=myurlcolor}
\usepackage{graphics,epstopdf,graphicx, amsthm, amsmath, amssymb, times, txfonts, braket, colortbl, color, bm}
\usepackage[up]{subfigure}
\usepackage{cleveref}

\definecolor{myurlcolor}{rgb}{0,0,0.7}

\theoremstyle{plain}

\providecommand{\theoremname}{Theorem}

\newcommand*{\myproofname}{Proof}

\makeatother

\begin{document}

\author{Uttam Singh}
\email{uttamsingh@hri.res.in}
\affiliation{Harish-Chandra Research Institute, Allahabad, 211019, India}
\affiliation{Homi Bhabha National Institute, Training School Complex, Anushakti Nagar, Mumbai 400085, India}
\author{Manabendra Nath Bera}
\email{manabendra.bera@icfo.es}
\affiliation{ICFO-The Institute of Photonic Science, Mediterranean Technology Park, 08860 Castelldefels (Barcelona), Spain}
\affiliation{Harish-Chandra Research Institute, Allahabad, 211019, India}
\author{Avijit Misra}
 \email{avijit@hri.res.in}
\affiliation{Harish-Chandra Research Institute, Allahabad, 211019, India}
\author{Arun Kumar Pati}
 \email{akpati@hri.res.in}
\affiliation{Harish-Chandra Research Institute, Allahabad, 211019, India}

\title{Erasing Quantum Coherence: An Operational Approach}

\begin{abstract}
Quantum coherence is one of the primary resources in quantum technologies and it plays a pivotal role in topical disciplines like quantum information, quantum thermodynamics, quantum biology to name a few. However, the resource theory of quantum coherence is very nascent and our understanding of quantum coherence is still limited from qualitative as well as quantitative perspectives. Towards this aim, here we provide an operational quantifier of the coherence of a quantum system in terms of the amount of noise that has to be injected into the system in order to fully decohere it. This quantifies the erasure cost of quantum coherence. We employ the entropy exchange between the system and the environment during the decohering operation and the memory required to store the information about the decohering operation as the quantifiers of noise. Both yield the same cost of erasing coherence in the asymptotic limit. In particular, we find that in the asymptotic limit, the minimum amount of noise that is 
required to fully decohere a quantum system, is equal to the relative entropy of coherence. This holds even if we allow for the nonzero small errors in the decohering process. As a consequence, it establishes that the relative entropy of coherence is endowed with an operational interpretation.
\end{abstract}
\maketitle
\section{Introduction}
\label{intro}
 With our ever increasing abilities to control systems at smaller and smaller scales, the quantum properties like quantum coherence and quantum entanglement make their presence felt more and more prominently. Recent developments in thermodynamics of nano scale systems suggest that the quantum coherence plays an essential role in determining the quantum state transformations and more importantly, in providing a family of second laws of thermodynamics \cite{Aspuru13, Horodecki2013, Skrzypczyk2014, Varun14, Brandao2015, Rudolph214, Rudolph114, Gardas2015, Avijit2015, Goold2015}. Also, the phenomenon of quantum coherence has been arguably attributed to the efficient functioning of some  complex biological systems \cite{Plenio2008, Arun2008, Aspuru2009, Lloyd2011, Li2012, Huelga13, Levi14}. Given the importance of quantum coherence, a formal structure of coherence resource theory is developed in recent years \cite{Gour2008, Marvian14, Baumgratz2014, Bromley2015, Girolami14, Marvian2014, Fan2014, 
Alex15, Fan15, Pinto2015, Du2015, Yao2015}. There are two inequivalent frameworks to characterize quantum coherence. The first framework is based on a set of incoherent operations as free operations and a set of freely available incoherent states \cite{Baumgratz2014}. This formalism has been successfully applied in the context of quantum entanglement, further providing a family of coherence monotones based on entanglement monotones \cite{Alex15} and quantification of the wave-particle duality \cite{Manab2015, Angelo2015}. The second formalism is based on the resource theory of asymmetry \cite{Gour2008, Marvian14, Marvian2014}, where operations are restricted to phase insensitive operations and symmetric states are free states \cite{Marvian14}. This formalism has been successfully employed in foundations of quantum thermodynamics \cite{Rudolph214, Rudolph114}. 

In the quantum information theory, to equip a particular ``resource'' of interest with an operational meaning, consideration of thermodynamic cost of destroying (erasing) the ``resource'', turns out to be very fruitful and far reaching \cite{Landauer1961, Bennett2003, Plenio2001, Oppenheim2002, Oppenheim03, Alicki2004, Rio2011}. For example, the Landauer erasure principle \cite{Landauer1961} has been a central one in laying the foundation of physics of information theory. Similarly, an operational definition of total correlation, classical correlation and quantum correlation is obtained independently in Refs. \cite{Groisman2005} and \cite{Horodecki2005}, considering the thermodynamic cost to erase the same. Additionally, it has been shown that the thermodynamic cost of erasing quantum correlation has to be associated with entropy production in the environment \cite{Pati2012}. This approach has also been successfully applied to private quantum decoupling \cite{Buscemi2009} and recently to markovianization \
cite{Wakakuwa2015}. Importantly, this approach can suitably be used for the quantification of the quantum resources \cite{Groisman2005}. In these tasks, quantum state randomization  \cite{Mosca2000, Boykin2003, Hayden2004} plays a pivotal role.

\begin{figure}
 \includegraphics[width=55 mm]{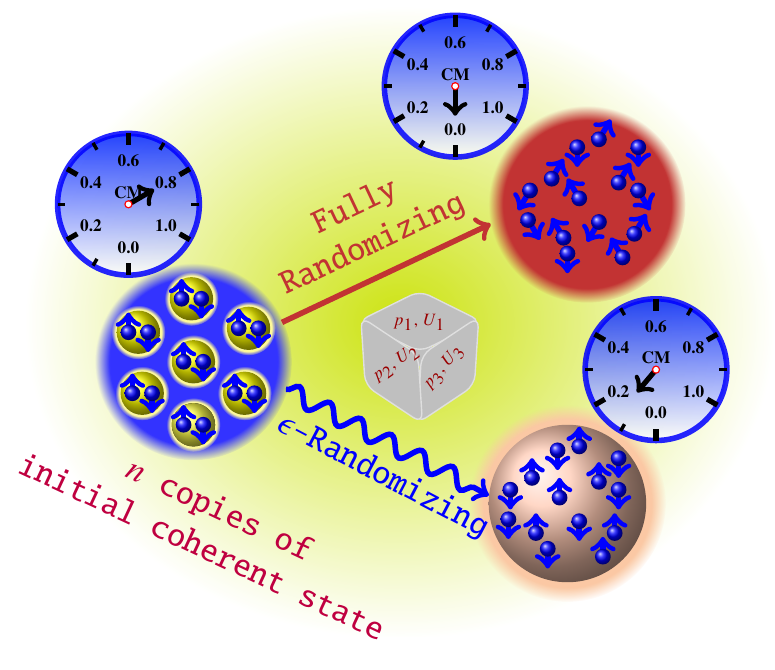}
 \caption{(Color online) Fully decohering and $\epsilon$-decohering maps: If we start with $n$ copies of any state (coherent or incoherent) and pass them through some decohering map, then the $n$ copies decohere completely if the map is fully decohering and if the map is $\epsilon$-decohering then the $n$ copies come very close to the fully decohered state keeping some amount of coherence which is close to zero. We show that in both the cases the minimum amount of noise that is required is same and is equal to relative entropy of coherence in the asymptotic limit.}\label{fig1}
\end{figure}

The resource theory of quantum coherence is still in its infancy as our understanding about it is limited from both qualitative and quantitative perspectives. Following the aforementioned operational approach, we quantify quantum coherence in terms of the amount of noise that has to be injected into the system such that the system decoheres completely. This, in turn, will provide operational meaning of the coherence.  We consider two different measures to quantify the amount of noise in the process of decohering a quantum system: first, the entropy exchange between system and environment during the decohering operation \cite{Schumacher1996, Schumacher96} and second, the memory required to store the information about the decohering operation \cite{Groisman2005}. We show that in the asymptotic limit, both these measures yields the same minimal cost of erasing coherence (the minimal noise required to fully decohere the system) and it turns out to be equal to the relative entropy of coherence \cite{
Baumgratz2014}. Relative entropy of coherence has been already identified as bona fide measure of coherence in the resource theory of coherence \cite{Baumgratz2014}, by considering the allowed operations to be incoherent operations and free states to be incoherent states. Here, to quantify coherence we followed a approach that is very much
different compared to the other measures existing in resource theory, in
the sense that we do not require our quantifier to follow the bona fide
criteria for it be a coherence monotone right from the beginning, but
interestingly this approach yields the same quantifier as the relative
entropy of coherence. Thus, our results provide an operational meaning of the relative entropy of coherence which in turn strengthens the basis of the coherence resource theory, in general. 

The paper is organized as follows: In Sec. \ref{sec:preliminaries}, we give a brief outline of the concepts required to understand the process of erasure of quantum coherence with illustrious examples. We present our main result of obtaining minimal cost of erasing coherence of a quantum system or of decohering a quantum system completely, in Sec. \ref{sec:erasure-cost}.
We conclude in Sec. \ref{sec:conclusion} with overview and implications of the results presented in the paper. In the Appendices, we provide some definitions such as the typical subspaces and present some useful theorems for the sake of completeness.

\section{Preliminaries: Various definitions}
\label{sec:preliminaries}
\noindent
{\it Quantum coherence:---} The theory of quantum coherence relies on fixed reference basis, as coherence is inherently a basis dependent quantity. For a given reference basis $\{\ket{a}\}$, the set of incoherent states $\mathcal{I}$ is defined as the set of all the states of the form $\rho_I = \sum_{a}p_a \ket{a}\bra{a}$, where $p_a \geq 0$, $\sum_a p_a=1$, and  the incoherent operations $\Lambda^\mathcal{I}$ are defined as completely positive trace preserving (CPTP) maps that transform the set of incoherent states onto itself. A function $C$ of density matrix $\rho$ is defined as a bona fide measure of coherence if it satisfies following conditions: (1) $C(\rho) =0$ iff $\rho \in \mathcal{I}$. (2) $C(\rho)$ is nonincreasing under the incoherent operations, i.e., $C(\Lambda^\mathcal{I}[\rho]) \leq C(\rho)$. (3) $C(\rho)$ is nonincreasing on an average under the selective incoherent operations, i.e., $\sum_i q_i C(\rho_i) \leq C(\rho)$, where $\rho_i = K_i \rho K_i^\dag/q_i$, $q_i = \mathrm{Tr}K_i \rho K_i^\dag$, and $K_i$ are the Kraus elements of an incoherent channel. (4) $C(\rho)$ is a convex function. One may note that the conditions (3) and (4) together imply the condition (2).

 The bona fide measures of coherence that emerge from this theory include the $l_1$ norm and the relative entropy of coherence \cite{Baumgratz2014}. The relative entropy of coherence of any state $\rho$ is defined as $C_{r}(\rho) = \min_{\sigma \in \mathcal{I}} D(\rho||\sigma)$, where $D(\rho||\sigma) = \mathrm{Tr} \rho (\log \rho - \log \sigma)$ is the relative entropy. After the minimization $C_{r}(\rho)$ is found to be equal to $H(\rho^d) - H(\rho)$, where $H(\rho)=-\mathrm{Tr} (\rho~ \mbox{ln}~ \rho)$, is the von Neumann entropy and $\rho^d=\sum_a \langle a | \rho |a \rangle | a \rangle\langle a |$ is the diagonal part of $\rho$ in the 
reference basis $\{\ket{a}\}$. Furthermore, the maximally coherent state is defined by $\ket{\psi_d} = \frac{1}{\sqrt{d}}\sum_{a=0}^{d-1} \ket{a}$, for which $C_r(\ket{\psi_d}\bra{\psi_d}) = \mbox{ln}~ d$. Moreover, a class of maximally coherent mixed states which satisfy a complementarity relation between coherence and mixedness, is proposed in Ref. \cite{Uttam2015} and is given by $\rho_p:=(1-p)\mathbb{I}_{d\times d}/d + p~ \ket{\psi_d}\bra{\psi_d}$, $0\leq p\leq 1$, for which $C_r(\rho_p) = \mbox{ln}~ d - H(\rho_p)$. The the $l_1$ norm of coherence is defined as $C_{l_1}(\rho) = \sum_{a\neq b} |\rho_{ab}|$, where $\rho = \sum_{a,b}\rho_{ab}\ket{a}\bra{b}$. The other measures of quantum coherence include skew information \cite{Girolami14} and geometric measures along with the entanglement based measures of coherence \cite{Alex15}.

Before we proceed further, we would like to give an illustration of a process of fully decohering a qubit quantum system in state $\ket{\psi_2} = \frac{1}{\sqrt{2}} \sum_{a=0}^{1}\ket{a}$, which is a maximally coherent state. Suppose we want to erase the coherence of this state. This can be achieved by applying two incoherent unitary transformations $\mathbb{I}_2$ and $\sigma_z$ with equal probability, i.e.,
\begin{align}
 \ket{\psi_2}\bra{\psi_2}\rightarrow \rho &= \frac{1}{2}\ket{\psi_2}\bra{\psi_2} + \frac{1}{2}\sigma_z\ket{\psi_2}\bra{\psi_2}\sigma_z =\frac{1}{2}\mathbb{I}_2.
\end{align}
Note that the final state is an incoherent state. This means that the application of two incoherent unitary operations with equal probability suffices to erase the coherence of the maximally coherent state. The same holds for the class of maximally coherent mixed states in two dimensions \cite{Uttam2015}. Also, it can be seen that for a $d$ dimensional quantum system, an ensemble of unitary transformations $\{\frac{1}{d^2}, \hat{X}^k\hat{Z}^j \}_{jk}$ exists, where $\hat X\ket{j}=\ket{j\oplus 1}, \hat Z\ket{j}=e^{\frac{2\pi ij}{d}}\ket{j}$ and $\oplus$ denotes addition modulo $d$, that can randomize any state $\rho$ of the system completely \cite{Hayden2004}, i.e.,
\begin{align}
\label{randomization-map}
 \rho \rightarrow \frac{1}{d^2}\sum_{j=1}^{d}\sum_{k=1}^{d} \hat{X}^k\hat{Z}^j \rho \hat{Z}^{j\dag}\hat{X}^{k\dag} =\frac{1}{d}\mathbb{I}_d.
\end{align}

But what is the cost to be paid in order to implement this probabilistic incoherent operation or how much noise does this operation inject into the system? One possibility, is to consider the amount of information (memory) needed to implement this (erasing) operation, which is related to the probabilities associated with the unitaries, and is equal to the Shanon entropy, $H(p=1/2) = 1$ bit for above qubit example. Therefore, one can say that applying an operation consisting of two elements, with equal probability, costs one bit of information or injects one bit of noise in the system. Similarly, for a qudit system, we can achieve the exact randomization via a map of the form Eq. (\ref{randomization-map}). The entropy that this map injects into the system as quantified by the amount of information needed to implement it, is given by $H(p=1/d^2) = 2\log_2 d$ bits. Clearly, the state independent randomization over-estimates the amount of noise that is necessary to decohere the state (cf. qubit and qudit cases). 
Also, this cost is independent of the nature of the operation, i.e., whether the operation is incoherent, unitary etc. The other choice to quantify the amount of noise injected into the system can be obtained based on exchange entropy as in Refs. \cite{Schumacher1996, Schumacher96, Groisman2005}. As we show below, the exchange entropy is smaller than $H(p)$.\\

\noindent {\it Exchange entropy:---} The exchange entropy \cite{Schumacher1996, Schumacher96} is defined as the amount of entropy that any channel $R$ injects into the system $S$ which passes through $R$. To define the exchange entropy, we purify the system state $\rho^S$ by a reference system $Z$ such that $\rho^S = \mathrm{Tr}_Z\ket{\psi}\bra{\psi}^{SZ}$. Now the entropy that $R$ injects into the system is defined as $H_{e} (R, \rho^S) := H \left((R \otimes  \mathbb{I}^Z) [\psi^{SZ}]\right)$, where $\mathbb{I}^Z$ is the identity operator on the reference system $Z$ and $H$ is the von Neumann entropy. The exchange entropy has been successfully employed in gaining insights in security of  cryptographic protocols \cite{Schumacher1996, Schumacher96}, in determining cost of erasing total, classical and quantum correlations \cite{Groisman2005}. Let $R$ be comprised of random unitary ensemble $\{p_i, U_i\}_{i=1}^{N}$. Then exchange entropy satisfies, $H_{e} (R, \rho^S) \leq H(p)\leq \log N$. For the example 
of maximally 
coherent qubit state, the entropy exchange is equal to one bit which is equal to the memory required to implement the erasing operation, as obtained in the preceding paragraph. Next we define general decohering map which can decohere any system and then $\epsilon$-decohering map that decoheres any state with small error $\epsilon > 0$.\\ 

\noindent
{\it Decohering and $\epsilon$-decohering maps:---} Let the decohering be achieved by an ensemble of incoherent unitaries $\{p_i, U_i^{I}\}_{i=1}^{N}$. We associate the map $\mathcal{R}:\rho \mapsto \sum_{i=1}^{N} p_i U^{I}_i\rho U_i^{I\dag}$, to the ensemble of these incoherent unitaries. We call this class of incoherent completely positive trace preserving (ICPTP) maps on system $S$ as the decohering maps. A decohering map $\mathcal{R}$ acting on a state $\rho$, is defined to be $\epsilon$-decohering map if there exists an incoherent state $\tau$ such that $||\mathcal{R}(\rho) - \tau||_1 \leq \epsilon$, where $||\cdot||_1$ is the trace norm \cite{Nielsen10, Wilde13} and for a matrix $A$, the trace norm is defined as $||A||_1 = \mathrm{Tr}\sqrt{A^\dag A}$. Note that the map $\mathcal{R}$ need not be comprised of incoherent unitaries. In fact, we show that a map $\mathcal{R}_c$, comprised of general unitaries, is equally suitable for decohering process. With these definitions in hand, 
we now proceed to present our results.

\section{Cost of erasing quantum coherence} 
\label{sec:erasure-cost}
We will mainly be concerned with the asymptotic case of the decohering procedure. But before going to the asymptotic case, let us consider the single copy scenario. Suppose a CPTP map $\Upsilon$ decoheres the system in any state $\rho$ and maps it to some incoherent state $\rho_I = \sum_{a}p_a\ket{a}\bra{a}$, where $\{\ket{a}\}$ is the fixed reference basis, $p_a \geq 0$ and $\sum_a p_a=1$, i.e., $\rho\rightarrow \Upsilon[\rho] = \rho_I$. The entropy exchange of this map is given by $H_e(\Upsilon,\rho) = H\left( (\Upsilon\otimes \mathbb{I}^Z)[\ket{\psi}\bra{\psi}^{SZ}]\right)$, where $Z$ is a reference system used to purify $\rho$. Now from monotonicity of the mutual information, i.e., $I( \Upsilon[\rho^{SZ}] ) \leq I(\rho^{SZ})$, we have 
\begin{equation}
 H_e(\Upsilon,\rho) \geq H(\rho_I) - H(\rho).
\end{equation}
The minimum exchange entropy is, $ H_e^{\min} = \min_{\{p_a\}}H(\rho_I) - H(\rho)$. Next, we will compute the $H_e^{\min}$ in the asymptotic limit when the CPTP map decoheres the state $\rho$ with some nonzero small error.

\smallskip
\noindent
{\bf Theorem:} The erasing cost of coherence of a quantum state $\rho$, as measured by the minimal amount of noise that has to be added in order to transform it into an incoherent state, is the relative entropy of coherence $C_{r}(\rho)$, in the asymptotic limit.
 Mathematically, 
\begin{align}
 &\sup_{\epsilon>0}\liminf_{n\rightarrow \infty}\frac{1}{n}\min \left\{H_e(\mathcal{R},\rho^{\otimes n}): \mathcal{R} ~~\epsilon \text{-decohering} \right\}\nonumber\\
 &=\sup_{\epsilon>0}\limsup_{n\rightarrow \infty}\frac{1}{n}\min \left\{\log N: \mathcal{R}_c ~~\epsilon \text{-decohering} \right\}
 =C_{r} (\rho),\nonumber
\end{align}
where $\mathcal{R}$ is defined as $\mathcal{R}:\rho \mapsto \sum_{i=1}^{N} p_i U^{I}_i\rho U_i^{I\dag}$ with $U_i^{I}$ being the incoherent unitary operator on the system and $\mathcal{R}_c$ is defined as $\mathcal{R}_c:\sigma\mapsto \frac{1}{N}\sum_{i=1}^{N} U_i \sigma U_i^{\dag}$ with $U_i$ being the unitary operator on the system. The relative entropy of coherence of any state $\rho$ is given by
\begin{equation}
 C_{r}(\rho) = H(\rho^d) - H(\rho),
\end{equation}
where $H(\rho)=-\mathrm{Tr} (\rho~ \mbox{ln}~ \rho)$, is the von Neumann entropy and $\rho^d=\sum_a \langle a | \rho |a \rangle | a \rangle\langle a |$ is the diagonal part of $\rho$ in the 
reference basis $\{\ket{a}\}$.

\smallskip
\noindent
{{\it Proof.---}The proof of the theorem follows from the following two lemmas.} 

\smallskip
\noindent
{\bf Lemma 1:} Consider an $\epsilon$-decohering map $\mathcal{R}$ on the $n$ copies of the system $S$ in the state $\rho$ as $\mathcal{R}:\rho^{\otimes n} \mapsto \sum_{i=1}^{N} p_i U^I_i\rho^{\otimes n} U_i^{I\dag}$, where $U^I_i$ is an incoherent unitary operator. Then, the amount of entropy that is injected into the system is lower bounded as $H_{e} (\mathcal{R}, \rho^{\otimes n})\geq n[C_r(\rho) - \epsilon \log d - H_{2}(\epsilon)]$, where $C_r(\rho)$ is the relative entropy of coherence for the state $\rho$ and $H_{2}(\epsilon) = -\epsilon \ln \epsilon - (1-\epsilon)\ln(1-\epsilon)$ is the binary Shanon entropy. In the asymptotic limit, the minimum entropy exchange, i.e., the minimum cost for erasing coherence, is given by
\begin{align}
\label{minimum-erasure-one}
 \sup_{\epsilon>0}\liminf_{n\rightarrow \infty}\frac{1}{n}\min \left\{H_e(\mathcal{R},\rho^{\otimes n}): \mathcal{R} ~~\epsilon \text{-decohering} \right\}=C_{r} (\rho).
\end{align}

\smallskip
\noindent
{\it Proof.---}First of all, define 
\begin{align}
 R_D := \mathcal{P}(\mathcal{R}[\rho^{\otimes n}]) = \sum_{\bm{k}}\Pi_{\bm{k}} \mathcal{R}[\rho^{\otimes n}]\Pi_{\bm{k}},
\end{align}
where $\{\Pi_{\bm{k}}\}$ are the projectors on the product subspaces written in the reference basis for the $n$ copies of the system. Any incoherent state under the projective measurement in the reference basis remains intact. Now utilizing the monotonicity of the trace norm under CPTP maps \cite{Nielsen10, Wilde13}, we have
\begin{align}
\label{RD}
 || R_D - \tau ||_1 &= || \mathcal{P}(\mathcal{R}[\rho^{\otimes n}]) - \mathcal{P}(\tau) ||_1\nonumber\\
 &\leq || \mathcal{R}[\rho^{\otimes n}] - \tau ||_1
 \leq \epsilon,
\end{align}
where in the last line we have used the fact that the map $\mathcal{R}$ is an $\epsilon$-decohering map. Now consider the following quantity
\begin{align}
 || \mathcal{R}[\rho^{\otimes n}] - R_D ||_1 &\leq || \mathcal{R}[\rho^{\otimes n}] - \tau||_1 + || \tau - R_D ||_1 \leq 2\epsilon,
\end{align}
where we have used the triangle inequality for the trace distance and made use of Eq. (\ref{RD}) together with the fact that the map $\mathcal{R}$ is an $\epsilon$-decohering map. Since $|| \mathcal{R}[\rho^{\otimes n}] - R_D ||_1 \leq 2\epsilon$, in the worst case one has $|| \mathcal{R}[\rho^{\otimes n}] - R_D ||_1 = 2\epsilon$. From the Fannes-Audenaert inequality \cite{Audenaert2007} (see also \cref{appendix:trace-distance}), we have
\begin{align}
 | H(\mathcal{R}[\rho^{\otimes n}]) - H(R_D) | &\leq \epsilon \ln (d^n-1) + H_{2}(\epsilon)\nonumber\\
 &\leq \epsilon n \log d + H_{2}(\epsilon),
\end{align}
where in the last line we have used $\ln (d^n-1) \leq n\log d$ and $H_{2}(\epsilon) = -\epsilon \ln \epsilon - (1-\epsilon)\ln(1-\epsilon)$. Noting the fact that $R_D$ is the diagonal part of $\mathcal{R}[\rho^{\otimes n}]$ and $H(R_D) \geq H(\mathcal{R}[\rho^{\otimes n}])$, we have
\begin{align}
\label{half}
 H(\mathcal{R}[\rho^{\otimes n}]) \geq H(R_D) - n\epsilon \log d - H_{2}(\epsilon).
\end{align}
Here, we pause to look at entropy of $R_D$ more closely. The incoherent unitary operations cannot change the diagonal parts of any density matrix except permuting the diagonal elements (of course they can change phases in off diagonal terms). This can be seen from the fact that any incoherent unitary $U^I$ can be written as a product of a unitary diagonal matrix $V$ and a permutation matrix $\Pi$, i.e., $U^I = V\Pi$. Therefore, we have $U^I\rho U^{I\dag} = V\sum_{ij}\rho_{ij}\ket{\Pi(i)}\bra{\Pi(j)} V^\dag$. In the following, a superscript $d$ on a state $\rho$ will mean the diagonal part of the density matrix in the fixed product reference basis. Now the diagonal part of the density matrix $U^I\rho U^{I\dag}$ is given by
\begin{align}
 (U^I\rho U^{I\dag})^d &= \sum_{l}\bra{l}V\sum_{ij}\rho_{ij}\ket{\Pi(i)}\bra{\Pi(j)} V^\dag\ket{l} \ket{l}\bra{l}\nonumber\\
 &= \sum_{i}\rho_{\Pi(i)\Pi(i)} \ket{\Pi(i)}\bra{\Pi(i)}.
\end{align}
Therefore, we have $H((U^I\rho U^{I\dag})^d) = H(\rho^d)$. Making use of this fact for $R_D$, we have
\begin{align}
 H(R_D) &\geq \sum_{i} p_i H \left( \Big(U^I_i \rho^{\otimes n} U_i^{I\dag}\Big)^d \right)\nonumber\\
 &= \sum_{i} p_i H \left(\rho^{d \otimes n}\right) = n H \left(\rho^d\right).
\end{align}
 From the Eq. (\ref{half}), we have
\begin{align}
 \label{half1p}
 H(\mathcal{R}[\rho^{\otimes n}]) &\geq n H(\rho^d) - n\epsilon \log d - H_{2}(\epsilon)\nonumber\\
 &\geq n [H(\rho^d) - \epsilon \log d - H_{2}(\epsilon)],
\end{align}
where in the last line, we have used $- H_{2}(\epsilon) \geq -n H_{2}(\epsilon) $. Now, we come to the question of finding the cost of decohering operation, i.e., the entropy that we have injected in the system. For this (as in the definition), we will consider the purification of $\rho$ which is given by $\psi$ such that $ \rho^{\otimes n} = \mathrm{Tr}_Z(\ket{\psi}\bra{\psi}^{\otimes n})$. Let us define
\begin{align}
 \Omega_{S^nZ^n} := (\mathbb{I}_Z^{\otimes n}\otimes \mathcal{R})[\ket{\psi}\bra{\psi}^{\otimes n}].
\end{align}
Since, $\mathcal{R}$ does not act on the reference system $Z$, $H(\Omega_{Z^n}) = H \left(\mathrm{Tr}_S(\ket{\psi}\bra{\psi}^{\otimes n})\right) = H(\rho^{\otimes n}) = n H(\rho)$. Now,
\begin{align}
 H_e(\mathcal{R},\rho^{\otimes n})=H(\Omega_{S^nZ^n}) &\geq H(\Omega_{S^n}) - H(\Omega_{Z^n})\nonumber\\
 &\geq H(\mathcal{R}[\rho^{\otimes n}]) - nH(\rho),
\end{align}
where in the first line, we have made use of the Araki-Lieb inequality \cite{Araki1970, Nielsen10, Wilde13}. Using Eq. (\ref{half1p}) in the above equation, we get
\begin{align}
 H_e(\mathcal{R},\rho^{\otimes n}) &\geq n [H(\rho^d)- H(\rho) - \epsilon \log d - H_{2}(\epsilon) ]\nonumber\\
 &= n [C_r(\rho) - \epsilon \log d - H_{2}(\epsilon)].
\end{align}
Therefore, in the asymptotic limit, the minimal entropy exchange is equal to the relative entropy of coherence as in Eq. (\ref{minimum-erasure-one})  (see Fig. \ref{fig1}). This completes the proof of the Lemma 1.

Next we consider the question of cost of erasing coherence while the amount of noise injected into the system is quantified by $\log N$, where $N$ is the number of unitaries in the ensemble comprising the $\epsilon$-decohering map.

\smallskip
\noindent
{\bf Lemma 2:} For any state $\rho$ and $\epsilon > 0$ there exists, for all sufficiently large $n$, a map $\mathcal{R}_c:\rho\mapsto \frac{1}{N}\sum_{i=1}^{N} U_i \rho U_i^{\dag}$ on system with $U_i$ being a unitary operator on the system, which $\epsilon$-decoheres it, and with $\log N \leq n\left(C_r(\rho) +\epsilon\right)$, where $C_r(\rho)$ is the relative entropy of coherence of the state $\rho$. In the asymptotic limit, the minimal amount of noise as quantified by $ \log N$, that is injected into the system is given by 
\begin{align}
\label{minimal-coherence-two}
 \sup_{\epsilon>0}\limsup_{n\rightarrow \infty}\frac{1}{n}\min \left\{\log N: \mathcal{R}_c ~~\epsilon \text{-decohering} \right\}=C_{r} (\rho).
\end{align}

\smallskip
\noindent
{\it Proof.---}Let us consider $n$ copies of the system in the state $\rho$. Also, consider the typical projector $\Pi$ that projects the system onto its typical subspace. Let $ \tilde{\rho} = \Pi \rho^{\otimes n} \Pi$. By definition of the typical projector, we have Tr$(\Pi\rho^{\otimes n} ) \geq (1-\epsilon)$. Therefore, using the ``gentle operator lemma'' \cite{Wilde13} (see also \cref{appendix:trace-distance}), we have
\begin{align}
 || \rho^{\otimes n} - \tilde\rho ||_1 \leq 2\sqrt{\epsilon}.
\end{align}
Now consider an ensemble of unitaries with some probability density function $p(dU)$, i.e. $\{U, p(dU)\}$ such that, for any state $\gamma$ on the typical subspace of $\rho^{\otimes n}$,
 $\int_{U} p(dU) U \gamma U^{\dag} = \frac{1}{D}\mathcal{I}_{\Pi}$,
where $D = 2^{n(H(\rho)-\epsilon)}$ and $\mathcal{I}_{\Pi}$ is the identity supported on the typical subspace of the system. Therefore, we have
\begin{align}
 \int_{U} p(dU) U \tilde\rho U^{\dag} = \frac{1}{D}\mathcal{I}_{\Pi}:=\tau \geq \frac{1}{D_d}\mathcal{I}_{\Pi},
\end{align}
where $D_d =2^{n(H(\rho^d)+\epsilon)}$.
Then, using the ``operator Chernoff bound'' \cite{Ahlswede2002, Ahlswede2003} (see also \cref{appendix:typical-subspace}), we show that we can select a subensemble of these unitaries which suffices the approximation. To this end, we consider 
$X:=D U \tilde\rho U^{\dag}$ as random operators with the distribution $p(dU)$.
Here $X\geq 0$. Using $\tilde\rho\leq \Pi/D$, we have
 $X= D U \tilde\rho U^{\dag} \leq U \Pi U^{\dag} \leq \mathcal{I}$.
Now, the average value $\mathbb{E}X$ of the random operator $X$ is given by
\begin{align}
 \mathbb{E}X &= D\int_{U} p(dU) U \tilde\rho U^{\dag}
 \geq \frac{D}{D_d}\mathcal{I}_{\Pi} = 2^{-n\left(C_r(\rho) +2\epsilon\right)}\Pi,
\end{align}
where $C_r(\rho)$ is the relative entropy of coherence of the state $\rho$. If $X_1,..,X_N$, where $X_i = D U_i \tilde\rho U_i^{\dag}$ ($i=1,..,n$), are $N$ independent realizations of $X$, then using the operator Chernoff bound, we have
\begin{align}
\label{Eq:sample-average}
 \text{Pr}&\left( (1-\epsilon)\mathbb{E}X \leq \frac{1}{N}\sum_{i=1}^{N}X_i \leq (1+\epsilon)\mathbb{E}X  \right)\nonumber\\
 &\geq 1 - 2 \text{~dim}(\Pi) \exp[-\frac{N\epsilon^2}{4\ln 2} 2^{-n\left(C_r(\rho) +2\epsilon\right)}].
\end{align}
For $N = 2^{n\left(C_r(\rho) +3\epsilon\right)} $ or higher, we have the corresponding probability on LHS of Eq. (\ref{Eq:sample-average}) nonzero for sufficiently large $n$. For this case, we have
 $(1-\epsilon)\mathbb{E}X \leq \frac{1}{N}\sum_{i=1}^{N}X_i \leq (1+\epsilon)\mathbb{E}X$.
This can be recast as
 $\left|\left| \frac{1}{N}\sum_{i=1}^{N} U_i \tilde\rho U_i^{\dag}  - \tau  \right|\right|_1 \leq \epsilon$.
Now, we have
\begin{align}
 \left|\left| \frac{1}{N}\sum_{i=1}^{N} U_i \rho^{\otimes n} U_i^{\dag}  - \tau  \right|\right|_1  
 &\leq \epsilon + || \rho^{\otimes n} -\tilde\rho||_1 \leq \epsilon + 2\sqrt{\epsilon}.
\end{align}
Therefore, there indeed exists decohering map $\mathcal{R}_c$ that $(\epsilon + 2\sqrt{\epsilon})$-decoheres any state  with, $N = 2^{n\left(C_r(\rho) +3\epsilon\right)} \leq 2^{n\left(C_r(\rho) +\epsilon + 2\sqrt{\epsilon}\right)}$, i.e., $\log N \leq n\left(C_r(\rho) +\epsilon + 2\sqrt{\epsilon}\right)$. Thus, in the asymptotic limit, the minimal cost of erasing coherence is given by Eq. (\ref{minimal-coherence-two}) (see Fig. \ref{fig1}). This concludes the proof of the Lemma 2.

Now, combining Lemmas 1 and 2, the proof of the theorem follows. It is worth emphasizing that we have not assumed any measure of coherence to start with, rather we have used two different quantifiers of the amount of noise, that are very important and well accepted in information theory (see also \cite{Groisman2005}). Rather, it is surprising that we find that the relative entropy of coherence emerges as the minimal amount of the noise that has to be added to the system to erase the coherence. Thus, our result provides an operational interpretation of the relative entropy of coherence developed in the resource theory of coherence \cite{Baumgratz2014}. Moreover, our result is robust, i.e., even if we allow for nonzero errors in the erasing process, we still get the same answer in the asymptotic limit.
Operational interpretations of various quantities in both classical and quantum information theory have been very striking with far-reaching impact on our understanding about the subject that has lead to explore different avenues. Operational interpretations of total and quantum correlations \cite{Groisman2005} are worth mentioning in this regard. However, computing the operational quantifiers is a formidable task in general. Thanks to the relative entropy of coherence that computing operational quantifier of coherence proposed by us is not a difficult task. Moreover, to the best of our knowledge, no other measure of coherence except the relative entropy of coherence and the coherence of formation \cite{Winter2015} is endowed with such an operational interpretation.

%

%

\section{Conclusion and Discussion}
\label{sec:conclusion}
To conclude, we have provided an operational quantifier of quantum coherence in terms of the amount of noise  that is to be  injected into a quantum system in order to fully decohere it. In the asymptotic limit, it is equal to the relative entropy of coherence. This provides the cost of erasing quantum coherence. It is worth mentioning that we have not assumed any of the measures of coherence to start with in order to prove our results. The relative entropy of coherence emerges naturally as the minimal erasing cost of coherence. Moreover, our result is robust, i.e., if we allow for nonzero error in the erasing process, it still gives the same answer in the asymptotic limit. In an independent work, Winter and Yang \cite{Winter2015} have shown that the relative entropy of coherence, emerges as the asymptotic rate at which one can distill maximally coherent states. This is very surprising as the same quantity, namely, the relative entropy of coherence comes up from two (apparently) completely 
different tasks such as the erasure and distillation of coherence. The resource theory of coherence starts with the premise that the allowed operations are the incoherent ones and the free states are the incoherent states, and thereby proposes the relative entropy of coherence as a valid measure of coherence along with the other measures like $l_1$ norm of coherence.
The formalism used in our work and in Ref. \cite{Winter2015} is well established and has far reaching implications in providing operational meaning to a resource, similar to other resources like quantum entanglement and correlations, in general.
In this regard, our results along with  results of Ref. \cite{Winter2015} further escalate the significance of the relative entropy of coherence as a bona fide measure of coherence.

In future one may ask the converse, i.e., in a complete protocol, what is the cost to keep a state coherent? Some partial 
results in a specific situation is provided in Ref. \cite{Vacanti2015}. It is worth proving that whether this cost is also equal to the relative entropy of coherence of $\rho$, in the asymptotic limit.
However, we leave it for future explorations. It is interesting to find a clear quantitative connection between our results and the Landauer's erasure principle \cite{Landauer1961} along with its improved and generalized versions \cite{Reeb2014, GooldA2015}.
Moreover, it will also be very interesting to further explore the quantitative relation between the no-hiding theorem \cite{Braunstein2007, Samal2011} and coherence erasure, both being very fundamental in their nature, as the no-hiding theorem applies to any process of hiding a quantum state, whether by randomization, thermalization or any other procedure. This will be the subject of future work. We hope that our results provide deep insights to the nature of coherence and interplay of information within the realm of quantum information and thermodynamics.  

\begin{acknowledgments}
We thank Andreas Winter for the useful remarks on the manuscript. We also thank Andreas Winter and Dong Yang for sharing their manuscript prior to submission on arXiv. U.S. and A.M. acknowledge the research fellowship of Department of Atomic Energy, Government of India. M.N.B. acknowledges funding from the Spanish project FOQUS (FIS2013-46768), 2014 SGR 874,
and the John Templeton Foundation.
\end{acknowledgments}

\appendix
\setcounter{equation}{0}
\section{The Fannes-Audenaert Inequality and the gentle operator lemma}
\label{appendix:trace-distance}

\smallskip{}
\noindent
{\it The Fannes-Audenaert inequality }:-- In the context of continuity of the von Neumann entropy, Audenaert proved a tighter inequality than the Fannes inequality \cite{Fannes1973}, which is now known as the Fannes-Audenaert inequality \cite{Audenaert2007} and can be stated as follows:
For any $\rho$ and $\sigma$ with $T\equiv\frac{1}{2}||\rho-\sigma||_{1}$, the following inequality holds
\begin{equation}
\left\vert H\left(  \rho\right)  -H\left(  \sigma\right)  \right\vert \leq T\log\left(  d-1\right)  +H_{2}(T),
\end{equation}
where $d$ is the dimension of the Hilbert space of the state $\rho$ and $H_{2}(T) = -T \ln T - (1-T)\ln (1-T) $ is the binary Shanon entropy. 

\smallskip{}
\noindent
{\it The gentle operator lemma} :-- The gentle operator lemma, which was first stated in Ref. \cite{Winter1999} and later improved in Ref. \cite{Ogawa2007} is stated as follows:
Suppose that a measurement operator $\Lambda$ ($0\leq\Lambda\leq I$) has a high probability of detecting a subnormalised state $\rho$, i.e., $\mathrm{Tr}\left\{  \Lambda\rho\right\}  \geq\mathrm{Tr}(\rho)-\epsilon$, where $1\geq\epsilon>0$ and $\epsilon$ is close to zero. Then $\sqrt{\Lambda}\rho\sqrt{\Lambda}$ is
close to the original state $\rho$ such that,
\begin{equation}
\left\Vert \rho-\sqrt{\Lambda}\rho\sqrt{\Lambda}\right\Vert _{1}\leq 2\sqrt{\epsilon},
\end{equation}
where $\left\Vert \sigma\right\Vert _{1} = \mathrm{Tr}{\sqrt{\sigma^\dag\sigma}}$.

\section{The typical subspaces and the operator Chernoff bound}
\label{appendix:typical-subspace}
\noindent
In this section, we give definitions of the typical subspaces and discuss their properties. See Ref. \cite{Wilde13} for further reading. 

\smallskip{}
\noindent
{\it Typical sequence and typical set}:-- Consider a sequence $x^{n}$ of $n$ realizations of a random variable $X$ which takes values $\{x\}$ according to probability distribution $\{p_X(x)\}$. A sequence $x^{n}$\ is $\delta$-\textit{typical} if its sample entropy $\overline{H}\left(  x^{n}\right) $, defined as $-\frac{1}{n}\log p_{X^n}(x^n)$, is $\delta$-close to the entropy $H\left(  X\right)  $\ of random variable $X$, where this random variable is the source of the sequence. The set of all $\delta$-typical sequences $x^n$ is defined as the typical set $T_{\delta}^{X^{n}}$, i.e.,
\begin{align}
 T_{\delta}^{X^{n}}\equiv\{x^n:|\overline{H}\left(  x^{n}\right) - H(X)|\leq \delta\}.
\end{align}
Now, consider a quantum state with spectral decomposition as 
\begin{align}
 \rho^X = \sum_{x} p_X(x) \ket{x}\bra{x}^X.
\end{align}
Considering $n$ copies of the state $\rho^X$, we have
\begin{align}
( \rho^X)^{\otimes n} :=  \rho^{X^{n}} = \sum_{x^n} p_{X^n}(x^n) \ket{x^n}\bra{x^n}^{X^{n}},
\end{align}
where $X^{n} = (X_1\ldots X_n)$, $x^{n} = (x_1\ldots x _n)$, $p_{X^n}(x^n) =p_{X}(x_1) \ldots p_{X}(x_n) $ and $\ket{x^n} = \ket{x_1}^{X_1}\otimes\ldots\otimes\ket{x_n}^{X_n}$.

\smallskip{}
\noindent
{\it Typical subspace}:-- The $\delta$-typical subspace $T_{\rho,\delta}^{X^{n}}$ is a subspace of the full Hilbert space $X_{1}$, \ldots, $X_{n}$ and is spanned by states $\left\vert
x^{n}\right\rangle ^{X^{n}}$\ whose corresponding classical sequences $x^{n}$\ are $\delta$-typical:%
\begin{equation}
T_{\rho,\delta}^{X^{n}}\equiv\text{\emph{span}}\left\{  \left\vert x^{n}\right\rangle ^{X^{n}}:x^{n}\in T_{\delta}^{X^{n}}\right\}.
\end{equation}
Also, one can define a typical projector, which projects a state onto the typical subspace, as
\begin{equation}
\Pi_{\rho, \delta}^{X^{n}}\equiv\sum_{x^{n}\in T_{\delta}^{X^{n}}}\left\vert x^{n}\right\rangle \left\langle x^{n}\right\vert ^{X^{n}}.
\end{equation}

\smallskip{}
\noindent
{\it Properties of typical subspaces}:-- 

\smallskip{}
\noindent
(a) The probability that the quantum state $\rho^{X^{n}}$ is in the typical subspace $T_{\rho, \delta}^{X^{n}}$ approaches one as $n$ becomes large:%
\begin{equation}
\forall\epsilon>0,\ \ \ \text{Tr}\left\{  \Pi_{\rho,\delta}^{X^{n}}\rho^{X^{n}}\right\}  \geq1-\epsilon,
\end{equation}
for sufficiently large $n$, 
where $\Pi_{\rho,\delta}^{X^{n}}$ is the typical subspace projector. 

\smallskip{}
\noindent
(b) The dimension $\dim\left(  T_{\rho, \delta}^{X^{n}}\right)  $\ of the $\delta$-typical subspace satisfies
\begin{equation}
\forall\epsilon>0,\ \ \ \left(  1-\epsilon\right)  2^{n\left(  H\left(  X\right)  -\delta\right)} \leq \text{Tr}\left\{  \Pi_{\delta}^{X^{n}}\right\}  \leq2^{n\left(  H\left( X\right)  +\delta\right)  },
\end{equation}
for sufficiently large $n$. 

\smallskip{}
\noindent
(c) For all $n$ the operator $\Pi_{\delta}^{X^{n}}\rho^{X^{n}}\Pi_{\delta}^{X^{n}}$ satisfies
\begin{equation}
\label{Eq:typ-state}
2^{-n\left(  H\left(  X\right)  +\delta\right)  }\Pi_{\delta}^{X^{n}}\leq \Pi_{\delta}^{X^{n}}\rho^{X^{n}}\Pi_{\delta}^{X^{n}}\leq2^{-n\left(  H\left( X\right)  -\delta\right)  }\Pi_{\delta}^{X^{n}}.
\end{equation}

\smallskip{}
\noindent
{\it The operator Chernoff bound}:-- 
Let $X_{1},\ldots,X_{n}$ $( \forall m\in\left[ n\right]  :0\leq X_{m}\leq I )$ be $n$ independent and identically distributed random operators with values in the algebra $\mathcal{B}\left(  \mathcal{H}\right)  $ of bounded linear operators on some Hilbert space $\mathcal{H}$. Let $\overline{X}$ denote the sample average of the $n$ random variables: $\overline{X}=\frac{1}{n}\sum_{m=1}^{n}X_{m}$. Suppose that for each operator $X_{m}$%
\begin{equation}
 \mathbb{E}_{X}\left\{  X_{m}\right\} \geq aI,
\end{equation}
where $a\in\left(  0,1\right) $ and $I$ is the identity operator on $\mathcal{H}$. Then for every $\epsilon$ where $0<\epsilon<1/2$ and $\left(  1+\epsilon\right)  a\leq1$, the probability that the sample average $\overline{X}$\ lies inside the operator interval $\left[  \left(  1\pm\epsilon\right)  \mathbb{E}_{X}\left\{  X_{m}\right\}\right]  $ is bounded as \cite{Ahlswede2002, Ahlswede2003},
\begin{align}
\label{Eq:op-Chernoff}
&\Pr_{X}\left\{  \left(  1-\epsilon\right)  \mathbb{E}_{X}  \left\{  X_{m}\right\}\leq\overline{X}\leq\left( 1+\epsilon\right)  \mathbb{E}_{X}\left\{  X_{m}\right\}\right\}\nonumber\\  &~~~~~~~~~~~~~~~~~~~~~~~~~~~\geq1-2\dim(\mathcal{H})\exp\left(  -\frac{n\epsilon^{2}a}{4\ln2}\right)  .
\end{align}

 \bibliographystyle{apsrev4-1}
 \bibliography{erase-lit}
 
\end{document}